\documentclass{article} 

\newcommand{\eprint}{\textsf} 

\newtoks\reportnoregister \newtoks\eprintnoregister
\newcommand{\reportnumber}[1]{\reportnoregister={#1}}
\newcommand{\eprintnumber}[1]{\eprintnoregister={#1}}

\reportnumber{\mbox{}} 
\eprintnumber{\mbox{}} 

\newcommand{\reportid}{
   \begin{minipage}{17cm}\vspace{-3.2cm}
     \begin{flushright}
      {\normalsize \the\reportnoregister \\[-.2cm]
            \eprint{\the\eprintnoregister}}\vspace{3.2cm}
     \end{flushright}
   \end{minipage}\hspace{-17cm} }

\catcode`@=11   
\def\title#1{\gdef\@title{\reportid#1}}
\catcode`@=12   
\newcommand{\gr}{04.20.-q}           
\newcommand{\symmcons}{11.30.-j}     

\usepackage{amsmath,amssymb}
\newcommand{\nc}{\newcommand}
\nc{\nms}{\negmedspace}
\nc{\nts}{\negthickspace}
\nc{\mcl}[1]{\mathcal{#1}}
\nc{\beq} {\begin{equation}}
\nc{\eeq} {\end{equation}}
\nc{\rarr} {\rightarrow}
\nc{\larr} {\leftarrow}
\nc{\lrarr} {\leftrightarrow}
\nc{\lbeq}[1]  {\label{eq: #1}}
\nc{\refeq}[1] {(\ref{eq: #1})}
\nc{\mrm}    {\mathrm}
\nc{\ga}{\Gamma}
\nc{\si}{\Sigma}
\nc{\back}{\!\!\!\!\!\!\!\!\!\!\!}
\nc{\nn}{\nonumber}
\nc{\mbf}[1] {{\mathbf #1}}
\nc{\bbet}{\bar{\beta}}
\nc\sgn{\mathop{\rm sgn}\nolimits}

\begin{document}

\reportnumber{USITP 98-14}
\eprintnumber{gr-qc/9807051}

\title{Third rank Killing tensors in general relativity. \\
       The (1+1)-dimensional case.}
\author{Max Karlovini\footnote{E-mail: \eprint{max@physto.se}}\; 
        and 
        Kjell Rosquist\footnote{E-mail: \eprint{kr@physto.se}} \\[10pt]
        {\small Department of Physics, Stockholm University}  \\
        {\small Box 6730, 113 85 Stockholm, Sweden} }
\date{}
\maketitle

\vspace{4cm}

\begin{abstract}{\normalsize
Third rank Killing tensors in (1+1)-dimensional geometries are
investigated and classified. It is found that a necessary and
sufficient condition for such a geometry to admit a third rank Killing 
tensor can always be formulated as a quadratic PDE, of order three or
lower, in a K\"ahler type potential for the metric. This is in
contrast to the case of first and second rank Killing tensors for
which the integrability condition is a linear PDE. The motivation for 
studying higher rank Killing tensors in (1+1)-geometries, is the fact
that exact solutions of the Einstein equations are often associated
with a first or second rank Killing tensor symmetry in the geodesic
flow formulation of the dynamics. This is in particular true for
the many models of interest for which this formulation is
(1+1)-dimensional, where just one additional constant of motion
suffices for complete integrability. We show that new exact solutions
can be found by classifying geometries admitting higher rank Killing
tensors.
}\end{abstract}

\vspace{1cm}
\centerline{\bigskip\noindent PACS numbers: \gr \quad \symmcons }
\clearpage


\section{Introduction}

Killing tensors are indispensable tools in the quest for exact
solutions in many branches of general relativity as well as classical
mechanics.  For nontrivial examples where Killing vectors ({\em i.e.}
first rank Killing tensors) and second rank Killing tensors have been
used to find and classify solutions of the Einstein equations the
reader is referred to \cite{rosquist:star,ru:kt,ujr:hh} and references
therein.  However, there are no examples of exact solutions which
correspond to third or higher rank Killing tensors.  Killing tensors
can also be important for solving the equations of motion in
particular spacetimes.  The notable example here is the Kerr metric
which admits a second rank Killing tensor \cite{pw:kt}.

However, none of the classical exact solutions of the Einstein
equations are known to admit higher rank Killing tensors.  Recently,
an example was given of a spacetime with a physically reasonable
energy-momentum tensor admitting a third rank Killing tensor
\cite{rg:intspacetimes}.  The method used in that work was based on
Lax pair tensors \cite{rosquist:lax}, a concept which can be viewed as
a generalization of Killing-Yano tensors \cite{dr:killing-yano}.  In
this paper we discuss the equations for third rank Killing tensors
using a more direct approach in the spirit of \cite{ru:kt} but
modified to take into account the qualitative differences in the third
rank case.  Our emphasis will be on ideas and concepts and most of the
results will be presented without proof.  The reader who wishes to see
more details can consult \cite{karlovini:killing}.

Any Killing tensor of rank two or higher has a traceless part which is
itself a conformal Killing tensor.  Furthermore the Killing tensor
equations (for rank two or higher) can be decomposed in a traceless
part and a trace part.  The traceless part constitutes the conformal
Killing tensor equations and involve only the traceless part of the
Killing tensor.  The trace part on the other hand involves both the
trace and the traceless parts.  In the second rank  case the equation
for the trace (which is then a scalar) gives rise to a covariant
integrability condition involving only the conformal Killing tensor.
In general such a covariant integrability condition is lacking for
Killing  tensors of rank three or higher.  However, since the trace
part of the Killing  tensor equations in the third rank case is itself 
a second rank tensor its  trace is a scalar equation.  It turns out
that this double trace equation is exactly  the condition that the
trace of the Killing tensor is divergence free.  The  third rank
Killing tensor equations therefore decompose into three parts, one
which involves only the conformal Killing tensor, one which involves
only the trace vector and finally one part which couples the trace to
the conformal Killing tensor.

In the present paper we focus on third rank Killing tensors in
(1+1)-dimensional geometries.  Such geometries are relevant to the
study of solutions of the Einstein equations for such diverse areas as
anisotropic cosmologies, inflationary cosmologies and relativistic
star models \cite{ujr:hh}. Applying our approach to the
(1+1)-dimensional case we are able to give a complete classification
of the third rank Killing tensors.  It turns out that any third rank
conformal Killing tensor can be uniquely characterized by a real
conformal Killing vector. This implies that there are two main types
of third rank Killing tensors depending on whether the causal
character of the conformal Killing vector is non-null or null. The
classification is then refined by considering the scalar product of
the conformal Killing vector with the trace vector. To solve the
Killing tensor equations the first step is to observe that the
divergence free property of the trace vector can be utilized to define
a scalar potential for the trace vector. Using the scalar potential
leads to simplification of the remaining Killing tensor equations. The
form of those equations depends on the causal character of the
conformal vector and on the scalar product of the conformal Killing
vector with the trace vector. However, in all cases it is
possible to find an integrability condition which involves only a
K\"ahler type potential for the metric. Unlike the second rank case
where the integrability conditions are linear, the third rank case
leads to integrability conditions which are quadratic in the K\"ahler
potential.

It has not been possible to find the general solution of the
integrability conditions, except in the case where both the conformal
Killing vector and the trace vector are null and have vanishing scalar 
product. However, we do give examples of solutions for all cases. We
also consider the special case where the metric admits a homothetic
Killing vector. In particular we give a complete treatment of the
homothetic metrics with two exponential terms. It turns out that the
only new integrable geometry in that case has complex exponential
coefficients and therefore has a trigonometric potential. It is in
fact a special case of a (1+1)-dimensional version of a 3-particle
Toda lattice. Except for the homothetic case, most of the solutions
given here represent new integrable (1+1)-dimensional geometries.

\section{Third rank Killing tensors in (1+1)-dimensional spacetimes}

Analogously to the second rank case investigated in \cite{ru:kt} we
shall make use of the fact that on any $n$-dimensional Riemannian or
Lorentzian manifold, a third rank Killing tensor can be decomposed
into its trace $K_\alpha$ and trace-free (conformal) part
$P_{\alpha\beta\gamma}$ according to
\begin{equation}
  K_{\alpha\beta\gamma} = P_{\alpha\beta\gamma} +
  \frac{3}{n+2}K_{(\alpha}g_{\beta\gamma)}.
\end{equation} 
This makes the Killing tensor equations
$K_{(\alpha\beta\gamma;\delta)}=0$ split into the conformal Killing
tensor equations for the trace-free part,  
\begin{equation}\lbeq{CKTeq}   
  C_{\alpha\beta\gamma\delta} := 
  P_{(\alpha\beta\gamma;\delta)} -
  \frac{3}{n+4}g_{(\alpha\beta}P^{\lambda}{}_{\gamma\delta);\lambda} =
  0,    
\end{equation}
and an equation which relates the trace-free part to the trace,
\begin{equation}\lbeq{coupling}
  D_{\alpha\beta} := K_{(\alpha;\beta)} +
  \frac{n+2}{n+4}P^\gamma{}_{\alpha\beta;\gamma} = 0. 
\end{equation}
By taking the trace of eq.\ \refeq{coupling}, one splits off the
condition that $K^\alpha$ be divergence-free, 
\begin{equation}\lbeq{divfree}
  K^\alpha{}_{;\alpha} = 0.
\end{equation}
Hence it is natural to start by solving the two decoupled conditions,
eq.\ \refeq{CKTeq} and \refeq{divfree}, before attempting to solve the
remaining (i.e. trace-free) part of eq.\ \refeq{coupling}. Focusing on
the (1+1)-dimensional case, we use null variables and write the
general metric as 
\begin{equation}\label{eq:metric}
  ds^2 = -2G(u,\bar{u})du d\bar{u}=-2\Omega^0\Omega^1,
\end{equation}
where we have introduced the standard null frame $\Omega^A$,
$A=0,\,1$, given by
\begin{equation}
  \Omega^0=G^{1/2}du, \quad \Omega^1=G^{1/2}d\bar{u}. 
\end{equation}
We shall consistently use the convention that the two-dimensional
tensor indices in this frame will take the values $0$ and $1$, while
in a coordinate frame they take the values $u$ and $\bar{u}$. To achieve
maximal simplication of the Killing tensor equations we use the following
parametrization of the Killing tensor (cf. the second rank case \cite{ru:kt})
\begin{equation}\lbeq{Killingparam}
  \begin{split}
    K_{000} &= -RG^{3/2} \\
    K_{111} &= -SG^{3/2} \\
    K_{001} &= -\textstyle\frac{1}{2}K_{u}G^{-1/2} \\
    K_{011} &= -\textstyle\frac{1}{2}K_{\bar{u}}G^{-1/2},
  \end{split}
\end{equation}
with $R := P^{\bar{u}\bar{u}\bar{u}}$ and $S := P^{uuu}$, using a
notation analogous to the second rank case. The difference, which is
solely due to the Killing tensor rank, being that $R$ and $S$
here are multiplied by $-G^{3/2}$ instead of $G$.

With the above parametrization, the conformal Killing tensor equations
are simply
\begin{equation}
  C_{0000} = -GR_{,u} = 0, \quad  C_{1111} = -GS_{,\bar{u}} = 0,
\end{equation}
requiring precisely that $R$ and $S$ be arbitrary functions of
$\bar{u}$ and $u$ respectively. This in fact shows that in two
dimensions, any third rank conformal Killing tensor
$P_{\alpha\beta\gamma}$ can in a unique way be represented by a
conformal Killing {\em vector} $\zeta_\alpha$ satisfying the equations
\begin{equation}
  C_{\alpha\beta} :=
  \zeta_{(\alpha;\beta)}-\frac{1}{n}\zeta^\gamma{}_{;\gamma}g_{\alpha\beta}=0.
\end{equation}
This can be shown as follows. In terms of the components $\zeta^u$ and
$\zeta^{\bar{u}}$, the conformal Killing vector equations in the
(1+1)-dimensional case reduce to
\begin{equation}
  C_{00} = -\zeta^{\bar{u}}{}_{,u}=0, \quad  C_{11} =
  -\zeta^{u}{}_{,\bar{u}}=0.
\end{equation}
These equations are solved by setting $\zeta^u=s(u)$,
$\zeta^{\bar{u}}=r(\bar{u})$. The existence of such a large class of
solutions reflects the fact that the conformal group in two dimensions
is of (uncountably) infinite dimension. By choosing $s(u)$ and
$r(\bar{u})$ appropriately, we can make our conformal Killing tensor
$P_{\alpha\beta\gamma}$ become the trace-free part of $\zeta_\alpha
\zeta_\beta \zeta_\gamma$, that is to say
\begin{equation}\lbeq{conformalreps}
  E_{\alpha\beta\gamma} := \zeta_\alpha\zeta_\beta\zeta_\gamma -
  \frac{3}{4}\zeta^\delta\zeta_\delta\zeta_{(\alpha}g_{\beta\gamma)} -
  P_{\alpha\beta\gamma} = 0.
\end{equation}
In component form, these equations become
\begin{equation}
  E_{000} = -\{[r(\bar{u})]^3-R(\bar{u})\}G^{3/2}=0, \quad
  E_{111} = -\{[s(u)]^3-S(u)\}G^{3/2}=0.
\end{equation}
Hence given any conformal Killing tensor $P_{\alpha\beta\gamma}$,
there is as claimed a unique {\em real} conformal Killing vector
$\zeta_\alpha$, given by $s(u)=[S(u)]^{1/3}$, $r(\bar{u}) =
[R(\bar{u})]^{1/3}$ (the real cubic roots), which represents
$P_{\alpha\beta\gamma}$ acoording to eq.\ \refeq{conformalreps}. We
shall use this result to characterize $P_{\alpha\beta\gamma}$
invariantly in terms of the causal character of $\zeta_\alpha$.

The divergence-free condition for $K^\alpha$ reads
\begin{equation}
  K^\alpha{}_{;\alpha} = -G^{-1}(K_{u,\bar{u}}+K_{\bar{u},u}) = 0,
\end{equation}
which we solve by setting $K_u = 2\Phi_{,u}$, $K_{\bar{u}} =
-2\Phi_{,\bar{u}}$ for some arbitrary potential function
$\Phi(u,\bar{u})$. This can be expressed covariantly in terms of the
natural volume form $\epsilon_{\alpha\beta} = G(du\wedge
d\bar{u})_{\alpha\beta}$ as
\begin{equation}
  K_\alpha = 2\epsilon_\alpha{}^\beta\Phi_{;\beta},
\end{equation}
making eq.\ \refeq{coupling} take the form
\begin{equation}\lbeq{Phi-coupling}
  D_{\alpha\beta} = 2\Phi_{;\gamma(\alpha}
  \epsilon_{\beta)}{}^\gamma + \frac23P^\gamma{}_{\alpha\beta;\gamma}
  = 0,
\end{equation}
The components of this equation, reading
\begin{align}\lbeq{D00}
  D_{00} &= 2\,(\frac{\Phi_{,u}}{G})_{,u} +
  \frac23G^{-2}(G^3R)_{,\bar{u}}=0, \\ \lbeq{D11}
  D_{11} &= -2\,(\frac{\Phi_{,\bar{u}}}{G})_{,\bar{u}} +
  \frac23G^{-2}(G^3S)_{,u}=0,
\end{align}
can be simplified by making a suitably chosen conformal transformation
$u=F(U)$, $\bar{u}=\bar{F}(\bar{U})$ together with a corresponding
frame scaling (boost) $\tilde{\Omega}^0=B\Omega^0$,
$\tilde{\Omega}^1=B^{-1}\Omega^1$, which up to the trivial
transformation $u\leftrightarrow \bar{u}$ will bring the Killing
tensor to one of three inequivalent standard forms. Since the
conformal factor transforms into $\tilde{G}=F'(U)\bar{F}'(\bar{U})G$,
the new frame will be defined analogously to the old one, but in new
null variables, by choosing $B=(\bar{F}'/F')^{1/2}$. We furthermore
write the inverse of the conformal transformation as $U=H(u)$,
$\bar{U}=\bar{H}(\bar{u})$. Now, a Killing tensor is called reducible
(and is thereby redundant for solving the geodesic equations) if it
can be written as a linear combination of symmetrized tensor products
of lower rank Killing tensors and the metric. Hence
$K_{\alpha\beta\gamma}$ is automatically reducible if the conformal
part $P_{\alpha\beta\gamma}$ is zero, since in that case the trace
$K_\alpha$ is required to satisfy the Killing vector equation.
Therefore we only take interest in the case when either $S(u)$ or
$R(\bar{u})$ is nonzero. Moreover if $S(u)$ and $R(\bar u)$ are both
nonzero (i.e. if $\zeta_\alpha$ is non-null) we can fix the conformal
gauge along the same lines as in \cite{ru:kt} by making a conformal
transformation which sets $S(u)$ and $R(\bar{u})$ to the standard
value $1$. However, in the case when $\zeta_\alpha$ is null so that
either $R$ or $S$ is zero, this requirement only fixes one of the new
variables $U$ and $\bar{U}$. To fix the other variable we use two
distinct conformal transformations depending on whether the scalar
product $\zeta^\alpha K_\alpha$ vanishes or not.  This makes it very
natural to define three major types of third rank Killing tensors
corresponding to the three qualitatively different ways in which the
conformal gauge is fixed. In table \ref{table:KT-type} this
classification is summarized invariantly in terms of the scalars
$\zeta^\alpha \zeta_\alpha$ and $\zeta^\alpha K_\alpha$.

For each Killing tensor type we shall now perform the conformal
transformation and derive a necessary and sufficient integrability
condition for eq.\ \refeq{Phi-coupling}. When doing this it will be
instructive to let the Killing tensor be represented in terms of the
geodesic invariant $I := K^{\alpha\beta\gamma}p_\alpha p_\beta
p_\gamma$, which has the general form
\begin{equation}
  I = Sp_u{}^3 + Rp_u{}^3 + 3(-\Phi_{,\bar{u}}p_u +
  \Phi_{,u}p_{\bar{u}})G^{-2}p_u p_{\bar{u}}. 
\end{equation}
\begin{table}[htbp]
  \begin{center}
  \begin{tabular}{|c|c|c|c|} \hline
    Killing tensor type & $\zeta^\alpha \zeta_\alpha$ & $\zeta^\alpha
    K_\alpha$ & Nontriviality condition \\ \hline
    I            & $\neq 0$ & no restriction & $K^\alpha K_\alpha\neq 0$,
    $\epsilon_{\alpha\beta}\zeta^\alpha K^\beta \neq 0$       \\ \hline 
    IIA          & $0$ & $\neq 0$ & $K^\alpha K_\alpha\neq 0$ \\ \hline
    IIB          & $0$ & $0$      &                            \\ \hline
  \end{tabular}
  \end{center}
\caption{Invariant classification of third rank Killing tensors in
  (1+1)-dimensional geometries.}
\label{table:KT-type}
\end{table}

\subsection*{Type I\,: $\zeta^\alpha \zeta_\alpha \neq 0.$}

Since an arbitrary conformal transformation brings $S(u)$ and
$R(\bar{u})$ into
\begin{align}
  \tilde{S}(U) &:= P^{UUU} = [H'(u)]^3P^{uuu} = [H'(u)]^3S(u), \\
  \tilde{R}(\bar{U}) &:= P^{\bar{U}\bar{U}\bar{U}} =
  [\bar{H}'(\bar{u})]^3P^{\bar{u}\bar{u}\bar{u}} =
  [\bar{H}'(\bar{u})]^3R(\bar{u}),
\end{align}
we obtain $\tilde{S}(U) = 1$, $\tilde{R}(\bar{U}) = 1$, by choosing
$H'(u)=[S(u)]^{-1/3}$, $\bar{H}'(\bar{u})=[R(\bar{u})]^{-1/3}$. With
this choice, eq.\ \refeq{D00} and \refeq{D11} are transformed into
\begin{align}\lbeq{D00I}
  \tilde{D}_{00} &:= B^{-2}D_{00} =
  2\,[(\frac{\Phi_{,U}}{\tilde{G}})_{,U} + \tilde{G}_{,\bar{U}}]=0, \\
  \lbeq{D11I}
  \tilde{D}_{11} &:= B^2 D_{11} =
  2\,[-(\frac{\Phi_{,\bar{U}}}{\tilde{G}})_{,\bar{U}}+\tilde{G}_{,U}]=0.
\end{align}
Evidently, if $\Phi_{,U}$ (or $\Phi_{,\bar{U}}$) is zero, $\bar{U}$
(or $U$) is required to be a null cyclic variable in $\tilde{G}$,
implying that the geometry is flat, and hence the case when
$\zeta_\alpha$ is non-null is interesting only when the trace
$K_\alpha$ is non-null as well. In table \ref{table:KT-type} this is
indicated as a nontriviality condition for type I.

The two equations \refeq{D00I} and \refeq{D11I} clearly have the
integrability condition
\begin{equation}
   (\frac{\Phi_{,U}}{\tilde{G}})_{,UU} +
   (\frac{\Phi_{,\bar{U}}}{\tilde{G}})_{,\bar{U}\bar{U}} = 0,  
\end{equation}
leading to
\begin{equation}\lbeq{dPhi}
\Phi_{,U} = -\tilde{G}\,\mathcal{K}_{,\bar{U}\bar{U}}, \quad
\Phi_{,\bar{U}} = \tilde{G}\,\mathcal{K}_{,UU}
\end{equation}
for some potential function $\mcl{K}$. Substituting this back into
eq.\ \refeq{D00I} and \refeq{D11I} yields
\begin{equation}
\begin{split}
  \tilde{D}_{00} &= 2(-\mcl{K}_{,U\bar{U}\bar{U}} +
  \tilde{G}_{,\bar{U}}) = 0, \\
  \tilde{D}_{11} &= 2(-\mcl{K}_{,UU\bar{U}} + \tilde{G}_{,U}) = 0,   
\end{split}
\end{equation}
showing that since $\mcl{K}$ is determined only up to
$\mcl{K}\rightarrow\mcl{K}+\psi$ with
$\psi_{,UU}=\psi_{,\bar{U}\bar{U}}=0$, it can be chosen such that
\begin{equation}\lbeq{Kpot}
  \tilde{G}=\mcl{K}_{,U\bar{U}}
\end{equation}
holds. Identifying our null variables with complex conjugate variables
and borrowing terminology from the theory of complex manifolds (see
e.g. \cite{nak}), the relation \refeq{Kpot} shows that $\mcl{K}$ plays
the role of a K\"{a}hler potential for the metric. A K\"{a}hler
potential $\mcl{K}$ has the property of transforming as a scalar under
conformal transformations since $G=\mcl{K}_{,u\bar{u}}$ clearly
implies $\tilde{G}=\mcl{K}_{,U\bar{U}}$. On the other hand it has the
disadvantage of being determined only up to a gauge transformation
$\mcl{K}\rightarrow\mcl{K}+f(u)+g(\bar{u})$. Requiring that $\mcl{K}$
should satisfy \refeq{dPhi} as well as \refeq{Kpot} however fixes the
gauge up to addition of a linear function of $U$ and $\bar{U}$.
Substituting eq.\ \refeq{Kpot} into eq.\ \refeq{dPhi} leads
immediately to the standardized integrability condition
\begin{equation}\lbeq{KcondI}
  (\mcl{K}_{,U\bar{U}}\mcl{K}_{,UU})_{,U} +
  (\mcl{K}_{,U\bar{U}}\mcl{K}_{,\bar{U}\bar{U}})_{,\bar{U}} = 0,
\end{equation}
which is necessary and sufficient for the existence of a third rank
Killing tensor of this type. Transforming back to the arbitrary 
null variables $u$ and $\bar{u}$ (without changing the K\"{a}hler
gauge), eq.\ \refeq{dPhi} becomes
\begin{equation}\lbeq{PhiKrelarb}
  \Phi_{,u} = -\mcl{K}_{,u\bar{u}}(R\mcl{K}_{,\bar{u}\bar{u}} + 
  \frac{1}{3}R'\mcl{K}_{,\bar{u}}),\quad  \Phi_{,\bar{u}} = 
  \mcl{K}_{,u\bar{u}}(S\mcl{K}_{,uu}+\frac{1}{3}S'\mcl{K}_{,u}), 
\end{equation} 
with the corresponding integrability condition
\begin{equation}\lbeq{KcondIarb}
  [\mcl{K}_{,u\bar{u}}(3S\mcl{K}_{,uu} + S'\mcl{K}_{,u})]_{,u} +
  [\mcl{K}_{,u\bar{u}}(3R\mcl{K}_{,\bar{u}\bar{u}} +
  R'\mcl{K}_{,\bar{u}})]_{,\bar{u}} = 0.
\end{equation}
Since the Killing tensor building blocks $P_{\alpha\beta\gamma}$,
$K_\alpha$ and $g_{\alpha\beta}$ all have been expressed in terms of
$S(u)$, $R(\bar{u})$ and $\mcl{K}$, we have the following closed
expression for the geodesic invariant:
\begin{equation}
  I = Sp_u{}^3 + Rp_{\bar{u}}{}^3 - [(3S\mcl{K}_{,uu} +
  S'\mcl{K}_{,u})p_u + (3R\mcl{K}_{,\bar{u}\bar{u}} +
  R'\mcl{K}_{,\bar{u}})p_{\bar{u}}]\mcl{K}_{,u\bar{u}}{}^{-1}p_u
  p_{\bar{u}}, 
\end{equation}
which in the standardized null variables simplifies to
\begin{equation}
  I = p_U{}^3 + p_{\bar{U}}{}^3 - 3(\mcl{K}_{,UU}\,p_U +
  \mcl{K}_{,\bar{U}\bar{U}}\,p_{\bar{U}})\mcl{K}_{,U\bar{U}}{}^{-1}p_U
  p_{\bar{U}}.
\end{equation}

\subsection*{Type II\,: $\zeta^\alpha \zeta_\alpha = 0.$}

Since the case $S(u)=0$ can be obtained from the case $R(\bar{u})=0$
by making the transformation $u\leftrightarrow \bar{u}$, we here only
need to consider the case $S(u)\neq 0$, $R(\bar{u}) = 0$. We then
solve eq.\ \refeq{D00} immediately by introducing a function
$Q(\bar{u})$ defined by the equation
\begin{equation}\lbeq{Qdef}
  \Phi_{,u} = Q(\bar{u})G.
\end{equation}
The choice of the transformation function $\bar{H}(\bar{u})$ will now
depend on whether $Q(\bar{u})$ is zero or non-zero, i.e. whether the
scalar product $\zeta^\alpha K_\alpha =
2(S^{1/3}\Phi_{,u}-R^{1/3}\Phi_{,\bar{u}})$ vanishes or not. However,
just as for type I we choose $H'(u)=[S(u)]^{-1/3}$ to obtain
$\tilde{S}(U)=1$, which regardless of $\bar{H}(\bar{u})$ makes eq.\
\refeq{D11} transform into 
\begin{equation}\lbeq{D11II}
  \tilde{D}_{11} =
  2\,[-(\frac{\Phi_{,\bar{U}}}{\tilde{G}})_{,\bar{U}}+\tilde{G}_{,U}]
  = 0.
\end{equation}
As for type I, $\Phi_{,\bar{U}}=0$ obviously leads to $U$ being a null 
cyclic variable in $\tilde{G}$ which implies a flat geometry. In
particular, this means that a Killing tensor of type IIA for which
$\zeta_\alpha$ is null and $\zeta^\alpha K_\alpha \neq 0$, can be
nontrivial only when the trace $K_\alpha$ is non-null. This is
indicated in table \ref{table:KT-type} as a nontriviality condition
for type IIA.

\subsection*{Type IIA\,: $\zeta^\alpha \zeta_\alpha = 0,$
  $\zeta^\alpha K_\alpha \neq 0.$} 

Since $Q(\bar{u})$ has the transformation property
\begin{equation}
  \tilde{Q}(\bar{U}) := \Phi_{,U}/\tilde{G} =
  \bar{H}'(\bar{u})\Phi_{,u}/G=\bar{H}'(\bar{u})Q(\bar{u})
\end{equation}
and here is nonzero, it is clear that the conformal gauge can be
fixed by choosing $\bar{H}'(\bar{u})=Q(\bar{u})^{-1}$ which makes
$\tilde{Q}(\bar{U})$ take the standard value $1$.

Working in the standardized null variables, we now substitute
$\tilde{G}=\Phi_{,U}$ into eq.\ \refeq{D11} to yield the final
nonlinear condition
\begin{equation}\lbeq{Phicond}
  \tilde{D}_{11} =
  2[-(\frac{\Phi_{,\bar{U}}}{\Phi_{,U}})_{,\bar{U}}+\Phi_{,UU}] = 0, 
\end{equation}
which in expanded form reads
\begin{equation}\lbeq{PhicondIIA}
  \Phi_{,U}{}^2\Phi_{,UU} +
  \Phi_{,\bar{U}}\Phi_{,U\bar{U}}-\Phi_{,U}\Phi_{,\bar{U}\bar{U}} = 0.
\end{equation}
Transforming back to arbitrary null variables, this condition becomes 
\begin{equation}\lbeq{PhicondIIAarb}
  \Phi_{,u}{}^2(S\Phi_{,uu} + \frac{1}{3}S'\Phi_{,u}) +
  Q^2(\Phi_{,\bar{u}}\Phi_{,u\bar{u}} -
  \Phi_{,u}\Phi_{,\bar{u}\bar{u}}) - QQ'\Phi_{,u}\Phi_{,\bar{u}} = 0. 
\end{equation}
Using now the fact that we have expressed $P_{\alpha\beta\gamma}$,
$K_\alpha$ and $g_{\alpha\beta}$ in terms of the functions $S(u)$,
$Q(\bar{u})$ and $\Phi$, the geodesic invariant takes the closed form
\begin{equation}
  I = Sp_u{}^3 + 3(-\Phi_{,\bar{u}}p_u +
  \Phi_{,u}p_{\bar{u}})Q^2\Phi_{,u}{}^{-2}p_u p_{\bar{u}}
\end{equation}
in arbitrary null variables, reducing to
\begin{equation}
   I = p_U{}^3 + 3(-\Phi_{,\bar{U}}p_U +
   \Phi_{,U}p_{\bar{U}})\Phi_{,U}{}^{-2}p_U p_{\bar{U}}
\end{equation}
in standardized null variables. If we let the metric be given by a
K\"{a}hler potential $\mcl{K}$ as $\tilde{G}=\mcl{K}_{,U\bar{U}}$, the 
relation $\tilde{G}=\Phi_{,U}$ shows that it is possible to make a
gauge transformation $\mcl{K}\rightarrow\mcl{K}+g(\bar{U})$ so that
$\mcl{K}_{,\bar{U}}=\Phi$ holds. Thus via eq.\ \refeq{Phicond},
$\mcl{K}$ is required to satisfy
\begin{equation}
  -(\frac{\mcl{K}_{,\bar{U}\bar{U}}}{\mcl{K}_{,U\bar{U}}})_{,\bar{U}}
  + \mcl{K}_{,UU\bar{U}}=0,
\end{equation}
leading directly to
\begin{equation}\lbeq{preKcond}
  \mcl{K}_{,U\bar{U}}\mcl{K}_{,UU} - \mcl{K}_{,\bar{U}\bar{U}} =
  h(U)\mcl{K}_{,U\bar{U}}
\end{equation}
for an arbitrary function $h(U)$. Using the remaining gauge freedom
for $\mathcal{K}$, the transformation $\mcl{K}\rightarrow\mcl{K}+f(U)$ 
with $f''(U)=h(U)$ makes eq.\ \refeq{preKcond} reduce to the
standardized form
\begin{equation}\lbeq{KcondIIA}
  \mcl{K}_{,U\bar{U}}\mcl{K}_{,UU} = \mcl{K}_{,\bar{U}\bar{U}},
\end{equation}
which corresponds to eq.\ \refeq{KcondI} for type I. Transforming back
to arbitrary null variables, the condition becomes
\begin{equation}\lbeq{KcondIIarb}
  \mcl{K}_{,u\bar{u}}(S\mcl{K}_{,uu} +
  \frac{1}{3}S'\mcl{K}_{,u}) =
  (Q\mcl{K}_{,\bar{u}})_{,\bar{u}}. 
\end{equation}
Finally we express the geodesic invariant in alternative form in terms 
of the gauge fixed K\"{a}hler potential instead of the trace
potential $\Phi$:
\begin{equation}
\begin{split}
  I &= Sp_u{}^3 - 3[(S\mcl{K}_{,uu} +
  \frac{1}{3}S'\mcl{K}_{,u})p_u -
  Qp_{\bar{u}}]\mcl{K}_{,u\bar{u}}{}^{-1}p_u p_{\bar{u}} \quad 
  \mathrm{(arbitrary\;null\;variables),} \\
    &= p_U{}^3 - 3(\mcl{K}_{,UU}\,p_U -
    p_{\bar{U}})\mcl{K}_{,U\bar{U}}{}^{-1}p_U p_{\bar{U}} \quad
    \mathrm{(standardized\;null\;variables).}
\end{split}
\end{equation}

\subsection*{Type IIB\,: $\zeta^\alpha \zeta_\alpha = 0,$
  $\zeta^\alpha K_\alpha = 0.$}

Since $Q(\bar{u})$ here vanishes, the scalar potential $\Phi$ is
according to eq.\ \refeq{Qdef} a function of $\bar{u}$ only and we
can hence introduce the function $P(\bar{u}):= \Phi_{,\bar{u}}$, which
transforms according to
\begin{equation}
  \tilde{P}(\bar{U}) := \Phi_{,\bar{U}} =
  [\bar{H}'(\bar{u})]^{-1}\Phi_{,\bar{u}} =
  [\bar{H}'(\bar{u})]^{-1}P(\bar{u}).
\end{equation}
If $P(\bar{u})$ vanishes, then so does the trace $K_\alpha$. Since
$\zeta_\alpha$ is null, this would mean that
\begin{equation}
  K_{\alpha\beta\gamma}=\zeta_\alpha \zeta_\beta \zeta_\gamma
\end{equation}
which implies that $\zeta_\alpha$ would be a null Killing
vector. Disregarding this trivial case, we see that the conformal
gauge can be fixed by making the choice $\bar{H}(\bar{u})=P(\bar{u})$
in order to obtain $\tilde{P}(\bar{U})=1$.

Substituting $\Phi_{,\bar{U}}=1$ into eq.\ \refeq{D11II} yields
\begin{equation}\lbeq{D11III}
  \tilde{D}_{11} = 2[-(\tilde{G}^{-1})_{,\bar{U}} + \tilde{G}_{,U}] =
  0,
\end{equation}
that is
\begin{equation}\lbeq{quasieq}
  \tilde{G}^2\tilde{G}_{,U}+\tilde{G}_{,\bar{U}}=0,
\end{equation}
which is a quasi-linear first order equation in the conformal factor
$\tilde{G}$. In arbitrary null variables this equation takes the form
\begin{equation}\lbeq{quasiIIBarb}
  G^2(SG_{,u}+\frac{1}{3}S'G)+PG_{,\bar{u}}-P'G=0.
\end{equation}
The geodesic invariant can here be directly expressed in terms of
$S(u)$, $P(\bar{u})$ and the conformal factor $G$ as
\begin{equation}
\begin{split}
  I &= Sp_u{}^3-3PG^{-2}p_u{}^2 p_{\bar{u}}\quad
  \mathrm{(arbitrary\;null\;variables)}, \\
    &= p_U{}^3-3\tilde{G}^{-2}p_U{}^2 p_{\bar{U}}\quad
    \mathrm{(standardized\;null\;variables).}
\end{split}
\end{equation}
To obtain a condition corresponding to eq.\ \refeq{KcondI} for type I
and eq.\ \refeq{KcondIIA} for type IIA, we substitute
$\tilde{G}=\mcl{K}_{,U\bar{U}}$ into eq.\ \refeq{D11III} to obtain
\begin{equation}
  -\mcl{K}_{,U\bar{U}}{}^{-1}+\mcl{K}_{,UU}=h(U),
\end{equation}
for an arbitrary function $h(U)$. Standardizing the condition, we let
$\mcl{K}\rightarrow\mcl{K}+f(U)$ with $f''(U)=h(U)$ yielding
\begin{equation}\lbeq{KcondIIB}
  \mcl{K}_{,U\bar{U}}\mcl{K}_{,UU}=1,
\end{equation}
or, in arbitrary null variables,
\begin{equation}
  \mcl{K}_{,u\bar{u}}(S\mcl{K}_{,uu}+\frac{1}{3}S'\mcl{K}_{,u}) = P. 
\end{equation}
As for type IIA, we now have an alternative expression for the geodesic
invariant, namely
\begin{equation}
\begin{split}
  I &= Sp_u{}^3 - (3S\mcl{K}_{,uu} +
  S'\mcl{K}_{,u})\mcl{K}_{,u\bar{u}}{}^{-1}p_u{}^2p_{\bar{u}} \quad
  \mathrm{(arbitrary\;null\;variables),} \\
    &= p_U{}^3 -
    3\mcl{K}_{,UU}\mcl{K}_{,U\bar{U}}{}^{-1}p_U{}^2p_{\bar{U}} \quad
    \mathrm{(standardized\;null\;variables)}.
\end{split}
\end{equation}

\subsection*{Comment on reducibility}

The number of independent Killing vectors can for a 2-dimensional
geometry be three, one or zero. The highly symmetric geometries that
admit three Killing vectors are precisely the ones that have constant
scalar curvature. Such geometries cannot have higher order
invariants that are independent of the three linear invariants. For
geometries with precisely one Killing vector $\xi_\alpha$ but no
irreducible second rank Killing tensors, a reducible third rank
Killing tensor can only be of the form
\begin{equation}\lbeq{K3red}
  K_{\alpha\beta\gamma} = C_1\xi_\alpha\xi_\beta\xi_\gamma +
  C_2\xi_{(\alpha}g_{\beta\gamma)}
\end{equation}
for some constants $C_1$ and $C_2$. The Killing vector $\xi_\alpha$ is 
by necessity non-null, since the geometry otherwise would be
flat. As we are not considering the automatically reducible case when
the trace-free part of a third rank Killing tensor vanishes, we assume
that $C_1\neq 0$ and redefine $K_{\alpha\beta\gamma}$ or $\xi_\alpha$
so that $C_1=1$. It then follows that the conformal Killing vector
$\zeta_\alpha$ coincides with $\xi_\alpha$ and that the trace
$K_\alpha$ is related to $\zeta_\alpha$ by 
\begin{equation}\lbeq{tracered}
  K_\alpha = (\zeta^\beta \zeta_\beta + \frac{4}{3}C_2)\zeta_\alpha.
\end{equation}
In particular, since $\zeta_\alpha$ is non-null, this reducible
Killing tensor is of type I. It would be practical to have an
invariant criterion which isolates this reducible case from the family
of type I Killing tensors since it cannot be identified by checking if 
the curvature is constant. In fact such a criterion does exist. Noting 
that a necessary and sufficient condition according to eq.\
\refeq{tracered} is that $\zeta_\alpha$ and $K_\alpha$ be parallell,
\begin{equation}\lbeq{parallell}
  \epsilon_{\alpha\beta}\zeta^\alpha K^\beta =
  2\zeta^\alpha\Phi_{;\alpha} = 0,
\end{equation}
we will now show that it is also a sufficient condition. This means
that given a third rank Killing tensor of type I, we must show that
eq.\ \refeq{parallell} implies that $\zeta_\alpha$ is a Killing vector
and that eq.\ \refeq{tracered} holds. Now, in the standard variables
for type I we have
\begin{equation}
  \zeta^\alpha = (\partial/\partial U + \partial/\partial
  \bar{U})^\alpha,
\end{equation}
so according to eq.\ \refeq{dPhi} and \refeq{parallell}, imposing that 
$\zeta_\alpha$ and $K_\alpha$ be parallell implies that 
\begin{equation}
  \zeta^\alpha\Phi_{;\alpha} = \Phi_{,U}+\Phi_{,\bar{U}} =
  \tilde{G}(-\mcl{K}_{,\bar{U}\bar{U}}+\mcl{K}_{,UU}) = 0,
\end{equation}
leading to $\mcl{K} = f(U+\bar{U}) + g(U-\bar{U})$. Substituting this
into the general integrability condition \refeq{KcondI} gives the
further restriction $f'''(U+\bar{U}) = 0$, i.e., up to irrelevant
linear terms in $\mcl{K}$,
\begin{equation}
\begin{split}
  \mcl{K}   &= \textstyle\frac{1}{2}A(U+\bar{U})^2 + g(U-\bar{U}) \\
  \tilde{G} &= A - g''(U-\bar{U})
\end{split}
\end{equation}
for some arbitrary constant $A$ and function $g(U-\bar{U})$. Clearly,
this shows that $\zeta_\alpha$ is a Killing vector. Furthermore eq.\
\refeq{dPhi} now implies that $K_\alpha$ can be written as $K_\alpha = 
(\zeta^\beta\zeta_\beta + 4A)\zeta_\alpha$. Comparing this with eq.\ 
\refeq{tracered} and reading off that $C_2=3A$ proves the assertion.

For geometries with an irreducible second rank Killing tensor, the
situation is different. If there are no Killing vectors, there are no
ways to construct a reducible third rank Killing tensor. This is of
course not the case if a Killing vector does exist, but for such
geometries we do not know of a simple invariant criterion which can be
used to check if a third rank Killing tensor is irreducible. 

To summarize, except for geometries which admit an irreducible second
rank Killing tensor and precisely one Killing vector, irreducibility
of a third rank Killing tensor is guaranteed if the geometry does not
have constant curvature and, for type I, if
$\epsilon_{\alpha\beta}\zeta^\alpha K^\beta \neq 0$.

\section{Some Solutions to the Standardized Integrability Conditions}

In this section we adress the problem of finding solutions to the
final integrability conditions expressed in the adapted null variables
$U$, $\bar{U}$. Due to the fact that these conditions are nonlinear
PDE's, in contrast to the corresponding conditions for the existence
of second rank Killing tensors \cite{ru:kt}, we shall have to settle
for giving some examples of nontrivial solutions, rather then giving
the general solutions. The exception is type IIB where the general
solution for the conformal factor can be given in implicit form.

\subsection*{Type I}

Let us begin by a remark on the symmetries of eq.\
\refeq{KcondI}. Obviously, the equation is invariant under coordinate
translations $U\rightarrow U+U_0$, $\bar{U}\rightarrow
\bar{U}+\bar{U}_0$ as well as under coordinate scalings $U\rightarrow
cU$, $\bar{U}\rightarrow c\bar{U}$ and scalings of the dependent
variable $\mcl{K}$. Moreover, the equation has a discrete $Z_3\times
Z_3$ symmetry of being invariant under $U\rarr e^{i2\pi m/3}U$,
$\bar{U}\rarr e^{-i2\pi n/3}\bar{U}$ with $m,\,n=0,\,\pm 1$. When writing
down explicit solutions below, we give only one representative in each
of these symmetry gauge classes.

Due to the scaling symmetries of eq.\ \refeq{KcondI}, it is natural to
make the ansatz that $\mcl{K}$ is a homogeneous function of $U$ and
$\bar{U}$, i.e.
\begin{equation}
  \mcl{K}(cU,c\bar{U})=c^{\lambda}\mcl{K}(U,\bar{U}).
\end{equation}
This implies that one can write $\mcl{K}=U^{\lambda}f(\eta)$ with
$\eta=\bar{U}/U$, which substituted into eq.\ \refeq{KcondI} yields a
complicated third order ODE for the function $f(\eta)$. For two values
of $\lambda$, namely $\lambda=1$ and $\lambda=2$, it is possible to
find the general solution to this equation. The solutions for these
two cases read 
\begin{equation}\lbeq{fokas}
  \left\{ \begin{array}{l} \mcl{K} =
      -U\!\int\!\int\!\eta^{-1/2}(\eta^3-1)^{-2/3}d\eta d\eta =
      \int\!\int\!\sqrt{U\bar{U}}(U^3 - \bar{U}^3)^{-2/3}dUd\bar{U}
      \\[5pt] 
           \tilde{G} = \sqrt{U\bar{U}}(U^3 - \bar{U}^3)^{-2/3},
           \\ \end{array}\right.
\end{equation}
\begin{equation}\lbeq{lambda2}
  \left\{ \begin{array}{l} \mcl{K} = U^2[A(\eta^{3/2} + 1)^{4/3} -
      B(\eta^{3/2}-1)^{4/3}] = A(U^{3/2}+\bar{U}^{3/2})^{4/3} -
      B(U^{3/2} - \bar{U}^{3/2})^{4/3},  \\[5pt]
           \tilde{G} = \sqrt{U\bar{U}}[A(U^{3/2} +
           \bar{U}^{3/2})^{-2/3} + B(U^{3/2} -
           \bar{U}^{3/2})^{-2/3}]. \\ \end{array}\right.
\end{equation}
The geometry corresponding to the solution \refeq{lambda2} is here
found to be superintegrable since it also admits a second rank
non-null Killing tensor. This can be shown by transforming into new
null variables $u=U^{3/2}$, $\bar{u}=\bar{U}^{3/2}$ after which the
conformal factor will satisfy the wave equation 
$G_{,uu}=G_{,\bar{u}\bar{u}}$ \cite{ru:kt}. Furthermore, for
$\lambda=3$ one has the special solution
\begin{equation}
  \left\{ \begin{array}{l} \mcl{K} = \frac{4}{9}U^3\eta^3 =
      \frac{4}{9}(U\bar{U})^{3/2}  \\[5pt] 
           \tilde{G} = \sqrt{U\bar{U}}, \\ \end{array}\right.
\end{equation}
which is trivial since it corresponds to a flat geometry. However, an
arbitrary linear combination of this solution and the solution
\refeq{fokas} also solves eq. \refeq{KcondI} and thus gives a
nontrivial generalization of the latter case. A lesson to be learnt
from this is that one should not reject homogeneous solutions which
are trivial as they stand since they are potential building blocks for 
nontrivial inhomogeneous solutions.

Introducing non-null variables $T$ and $X$ defined by $U=T+X$,
$\bar{U}=T-X$ in terms of which
\begin{equation}
  \tilde{G}=\frac{1}{4}(\mcl{K}_{,TT}-\mcl{K}_{,XX}),
\end{equation}
one easily verifies that eq.\ \refeq{KcondI} is solved by letting
$\mcl{K}$ be an arbitrary function of $X=(U-\bar{U})/2$
only, which corresponds to $\zeta_\alpha$ being a Killing
vector. Consequently, the equation also has the complex solutions when 
$\mcl{K}$ is a function of $X\pm i\sqrt{3}T = -e^{\mp i2\pi/3}U+e^{\pm
  i2\pi/3}\bar{U}$ only. These are of course trivial solutions by
themselves, but they suggest that the ansatz
\begin{equation}
  \mcl{K} = f(-2X)+g(X+i\sqrt{3}T)+h(X-i\sqrt{3}T)
\end{equation}
be made, for the simple reason that each term by itself satisfies the
equation. Moreover, we shall assume that the three functions have the
same functional dependence, i.e. that $f(z)=g(z)=h(z)$, thus ensuring
that $\mcl{K}$ is real and invariant under the $Z_3$ symmetry $U\rarr
e^{i2\pi n/3}U$, $\bar{U}\rarr e^{-i2\pi n/3}\bar{U}$. Some special
solutions obtained with this ansatz are
\begin{equation}\lbeq{toda}
  \left\{ \begin{array}{l} f(z) = -e^z-\frac{1}{6}Az^2 \\[5pt]
           \tilde{G} = e^{-2X} + e^{X+i\sqrt{3}T} + e^{X-i\sqrt{3}T} + 
           A = e^{-2X} + 2e^{X}\cos{\sqrt{3}T} + A, \\
         \end{array}\right.
\end{equation}
\begin{equation}\lbeq{constcurv}
  \left\{ \begin{array}{l} f(z) = \ln{z} - \frac{1}{6}Az^2
      \\[5pt]
          \begin{split}
           \tilde{G} &=
           \displaystyle (-2X)^{-2} + (X+i\sqrt{3}T)^{-2} +
           (X-i\sqrt{3}T)^{-2} + A \\[5pt] 
                     &= \frac{9(X^2-T^2)^2}{4X^2(X^2+3T^2)^2} + A, \\
          \end{split}
          \end{array}\right.
\end{equation}
\begin{equation}
  \left\{ \begin{array}{l} f(z) = \ln{\!(\sinh{z})} - \frac{1}{6}Az^2
      \\[5pt]
          \begin{split} 
           \tilde{G} &= \displaystyle [\sinh{\!(-2X)}]^{-2} +
           [\sinh{\!(X+i\sqrt{3}T)}]^{-2} + [\sinh{\!(X-i\sqrt{3}T)}]^{-2} + A 
           \\[5pt]
                     &= 2\,\frac{\cos{4\sqrt{3}T} +
             2(\cosh{6X} - 3\cosh{2X})\cos{2\sqrt{3}T} - 3\cosh{4X} +
             6}{(2\sinh{2X}\cos{2\sqrt{3}T} - \sinh{4X})^2} + A, \\
          \end{split}
         \end{array}\right.
\end{equation}
\begin{equation}\lbeq{sinsep}
  \left\{ \begin{array}{l} f(z) = \ln{\!(\sin{z})} - \frac{1}{6}Az^2
      \\[5pt]
          \begin{split}
           \tilde{G} &= \displaystyle [\sin{\!(-2X)}]^{-2} +
           [\sin{\!(X+i\sqrt{3}T)}]^{-2} + [\sin{\!(X-i\sqrt{3}T)}]^{-2} + A
           \\[5pt]
                     &= 2\,\frac{\cosh{4\sqrt{3}T} + 2(\cos{6X} -
             3\cos{2X})\cosh{2\sqrt{3}T} - 3\cos{4X} +
             6}{(2\sin{2X}\cosh{2\sqrt{3}T} - \sin{4X})^2} + A. \\
          \end{split}
         \end{array}\right. 
\end{equation}
All of the solutions \refeq{toda} - \refeq{sinsep} correspond to
well-known classical mechanical potentials that are integrable with a
cubic invariant \cite{hiet}. In particular, the conformal factor for
the first solution \refeq{toda} is the Lorentzian analogue of the
three-particle Toda potential. This solution differs from the others
obtained with the given ansatz in that its three exponential terms can
have arbitrary constant coefficients, which means for this case it is
not necessary that $\mcl{K}$ obeys the $Z_3$ symmetry. In the case
when the arbitrary constant $A$ is zero, the metric corresponding to
the solution \refeq{constcurv} has constant but nonzero curvature and
then admits three independent non-null Killing vectors.

When setting $\mcl{K}$ to an arbitrary function of $T$ only, eq.\
\refeq{KcondI} requires that this function be a second degree
polynomial. A natural ansatz is therefore obtained by replacing the
polynomial coefficients with arbitrary functions of $X$, i.e. 
\begin{equation}
  \mcl{K}=f(X)T^2+g(X)T+h(X).
\end{equation}
The general solution with this ansatz reads
\begin{equation}
  \left\{ \begin{array}{l} \mcl{K} =
      (-9AX^{4/3} + 4D)T^2 - 9BX^{4/3}T - \frac{27}{14}AX^{10/3} -
      9CX^{4/3} + 2DX^2 \\[5pt] 
  \tilde{G} = -\frac{3}{4}AX^{4/3} + (AT^2 + BT + C)X^{-2/3} + D, \\ 
\end{array}\right.
\end{equation}
with some irrelevant integration constants set to zero. In the case
when $A\neq 0$, the solution can be further standardized by setting
$A=1$, $B=0$. It can then be identified as the Lorentzian analogue of
Holt's integrable classical mechanical potential \cite{hiet}. If $A=0$
but $B\neq 0$, we still have a nontrivial solution which is
standardized by setting $B=1$, $C=0$.

We have seen that imposing that $\zeta_\alpha$ and $K_\alpha$ be
parallell leads to $\zeta_\alpha$ being a Killing vector in terms of
which $K_{\alpha\beta\gamma}$ is reducible. We here instead make the
ansatz that $\zeta_\alpha$ and $K_\alpha$ be orthogonal,
\begin{equation}
  \zeta^\alpha K_\alpha = 2(\Phi_{,U}-\Phi_{,\bar{U}}) = -
  2\mcl{K}_{,U\bar{U}}(\mcl{K}_{,\bar{U}\bar{U}} + \mcl{K}_{,UU}) = 0.
\end{equation}
This leads directly to $\mcl{K}$ being a harmonic function of $U$ and
$\bar{U}$, i.e.
\begin{equation}\lbeq{Kanalytic}
  \mcl{K} = f(U+i\bar{U})+\bar{f}(U-i\bar{U}),
\end{equation}
where $f$ is an analytic function of $U+i\bar{U}$. When substituting
eq.\ \refeq{Kanalytic} into eq.\ \refeq{KcondI} one obtains the
nontrivial solution
\begin{equation}
  \left\{ \begin{array}{l} \mcl{K} =
      -\frac{2}{15}\{[(1+i)(U+i\bar{U})]^{5/2} +
      [(1-i)(U-i\bar{U})]^{5/2}\} \\[5pt]  
  \tilde{G} = \sqrt{(1+i)(U+i\bar{U})} + \sqrt{(1-i)(U-i\bar{U})} =
  \sqrt{2[\sqrt{2(U^2+\bar{U}^2)}+U-\bar{U}]}. \\
\end{array}\right.
\end{equation}
Since the conformal factor satisfies the Laplace equation
$\tilde{G}_{,UU}+\tilde{G}_{,\bar{U}\bar{U}}=0$, the geometry also 
admits a non-null second rank Killing tensor \cite{ru:kt} and is thus
superintegrable.

\subsection*{Type IIA}

We shall here give the results in terms of the trace potential $\Phi$,
which in this case also serves as a potential for the metric via the
relation $\tilde{G}=\Phi_{,U}$. Contrary to the K\"{a}hler potential
condition \refeq{KcondIIA}, the equivalent condition
\refeq{PhicondIIA} imposed on $\Phi$ is a PDE which is linear in the
second derivatives. However, this does not by necessity mean that eq.\
\refeq{PhicondIIA} in general is easier to work with than eq.\
\refeq{KcondIIA}, since the latter has the advantage of being
quadratic instead of cubic in the dependent function, besides being a
more compact equation.

Noting that eq.\ \refeq{PhicondIIA} is invariant under translations
$U\rarr U+U_0$, $\bar{U}\rarr \bar{U}+\bar{U}_0$ as well as the
correlated scalings $U\rarr aU$, $\bar{U}\rarr b\bar{U}$, $\Phi\rarr
a^{-3}b^2\Phi$ (that is to say, if $\Phi=f(U,\bar{U})$ is a solution,
then so is $\Phi=a^{-3}b^2f(aU,b\bar{U})$\,), the solutions can
preferably be exhibited with the freedom to make these transformations
fixed, but it is often convenient to avoid a complete fixing in
order to be able to let several inequivalent subcases be contained in
one single expression.

By trial and error, one quickly finds that the two ans\"{a}tze
\begin{align}
  \Phi &= f(\bar{U})U^3+g(\bar{U})U^2+h(\bar{U})U+k(\bar{U}), \\
  \Phi &= f(\bar{U})U^3+g(\bar{U})U^{3/2}+h(\bar{U}),
\end{align}
give rise to nontrivial solutions obtained by solving ODE's for the
coefficient functions. The following solutions have been found:
\begin{equation}
  \left\{ \begin{array}{l} \Phi = \frac{1}{9}\bar{U}^{-2}U^3 +
      (A\bar{U}^{-2} + B\bar{U}^{-1/3})U^2 + [3(A\bar{U}^{-1} +
      B\bar{U}^{2/3})^2 + C\bar{U}^{-2/3}]U \\[5pt]
      \qquad + 3(A\bar{U}^{-2/3} + B\bar{U})^3 + 3C(A\bar{U}^{-2/3} +
      B\bar{U})  \\[5pt]
      \tilde{G} = \frac{1}{3}\bar{U}^{-2}U^2 + 2(A\bar{U}^{-2} +
      B\bar{U}^{-1/3})U + 3(A\bar{U}^{-1} + B\bar{U}^{2/3})^2 +
      C\bar{U}^{-2/3} \\ \end{array}\right.
\end{equation}
\begin{equation}
  \left\{ \begin{array}{l} \Phi =
      -\frac{1}{9}(\cosh{\bar{U}})^{-2}U^3 + A(\cosh{\bar{U}})^{-2/3}U 
      \\[5pt] 
           \tilde{G} = - \frac{1}{3}(\cosh{\bar{U}})^{-2}U^2 +
           A(\cosh{\bar{U}})^{-2/3}  \\ \end{array}\right. 
\end{equation}
\begin{equation}
  \left\{ \begin{array}{l} \Phi = \frac{1}{9}(\cos{\bar{U}})^{-2}U^3 + 
      A(\cos{\bar{U}})^{-2/3}U  \\[5pt] 
           \tilde{G} = \frac{1}{3}(\cos{\bar{U}})^{-2}U^2 +
           A(\cos{\bar{U}})^{-2/3}  \\ \end{array}\right. 
\end{equation}
\begin{equation}\lbeq{K2A}
  \left\{ \begin{array}{l} \Phi = \frac{1}{2}AU^2 +
      (\frac{1}{2}A^2\bar{U}^2 + B\bar{U}+C)U  \\[5pt] 
\qquad + \frac{1}{8}A^3\bar{U}^4 + \frac{1}{2}AB\bar{U}^3 +
\frac{1}{2}(A^{-1}B^2+AC)\bar{U}^2 + A^{-1}BC\bar{U}  \\[5pt]
           \tilde{G} = AU + \frac{1}{2}A^2\bar{U}^2 + B\bar{U} + C  \\
         \end{array}\right. 
\end{equation}
\begin{equation}\lbeq{K2B}
  \left\{ \begin{array}{l} \Phi = \frac{1}{2}Ae^{2\bar{U}}U^2 + (
      \frac{1}{6}Ae^{4\bar{U}} + Be^{2\bar{U}} + Ce^{\bar{U}}  )U
      \\[5pt]  
  \qquad + \frac{1}{6}A(\frac{1}{12}A^2e^{6\bar{U}} + Be^{4\bar{U}} +
  Ce^{3\bar{U}}) + A^{-1}B(\frac{1}{2}Be^{2\bar{U}} + Ce^{\bar{U}})
  \\[5pt] 
           \tilde{G} = Ae^{2\bar{U}}U + \frac{1}{6}A^2e^{4\bar{U}} +
           Be^{2\bar{U}} + Ce^{\bar{U}}  \\ \end{array}\right.
\end{equation}
\begin{equation}
  \left\{ \begin{array}{l} \Phi = \frac{1}{9}\bar{U}^{-2}U^3 +
      \frac{2}{3}(A\bar{U}^{-1/2} + B\bar{U}^{-3/2})U^{3/2} +
      A^2\bar{U} + B^2\bar{U}^{-1}  \\[5pt] 
           \tilde{G} = \frac{1}{3}\bar{U}^{-2}U^2 + (A\bar{U}^{-1/2} + 
           B\bar{U}^{-3/2})U^{1/2}  \\ \end{array}\right. 
\end{equation}
According to \cite{ru:kt}, $\tilde{G}_{,UU}=0$ is up to $U\lrarr
\bar{U}$ the standardized integrability condition for the existence of 
a second rank Killing tensor with a null eigenvector. Hence the two
solutions \refeq{K2A} and \refeq{K2B} correspond to superintegrable
geometries, admitting both second and third rank Killing tensors.

\subsection*{Type IIB}

For this class of third rank Killing tensors it is possible to write
down the implicit general solution to the quasi-linear, first order
condition \refeq{quasieq} as
\begin{equation}\lbeq{implicitG}
  F(\xi,\eta)=0,
\end{equation}
where $F$ is an arbitrary function of its two arguments $\xi :=
U-\tilde{G}^2\bar{U}$ and $\eta := \tilde{G}$. Symmetries ensure
that if $\tilde{G} = f(U,\bar{U})$ is a solution, then so is
$\tilde{G} = f(U+U_0,\bar{U}+\bar{U}_0)$, $\tilde{G} =
\sqrt{b/a}f(aU,b\bar{U})$ and $\tilde{G} = [f(\bar{U},U)]^{-1}$. A few
explicit solutions can be obtained by choosing $F$ such that eq.\
\refeq{implicitG} becomes a polynomial equation in $\tilde{G}$ of
sufficiently low order. The simplest nontrivial example of such a
solution is obtained by setting $F(\xi,\eta)=\xi-2\eta$. Solving the
corresponding second order equation in $\tilde{G}$ yields
\begin{equation}
  \tilde{G}=\left(1\pm\sqrt{1+U\bar{U}}\right)\bar{U}^{-1}.
\end{equation}

\section{Geometries admitting a homothetic vector field}

In this section we consider the class of (1+1)-dimensional geometries
that admit a homothetic vector field $\xi$, satisfying $\mcl{L}_\xi
g_{\alpha\beta} = 2g_{\alpha\beta}$. In the physical applications we
have in mind the homothetic vector is timelike and we therefore
restrict attention to this case.  In fact, this implies no loss of
generality since the timelike and spacelike cases are mathematically
equivalent and the lightlike case is uninteresting as it requires a
flat geometry. Adapting the coordinates to $\xi$, the metric can be
written in the form
\begin{equation}
  ds^2=2e^{2t}F(x)(-dt^2+dx^2),
\end{equation}
with $\xi=\partial/\partial t$. Referring to (\ref{eq:metric}) and
introducing the null variables $u=t+x$, $\bar{u}=t-x$ we see that
$G=e^{2t}F(x)$ and that the corresponding null frame $\Omega^A$ is
given by $\Omega^0=e^tF^{1/2}(dt+dx)$, $\Omega^1=e^tF^{1/2}(dt-dx)$. 
We assume that the Killing tensor, like the metric, has the rescaling
property
\begin{equation}
  \mcl{L}_\xi K_{\alpha\beta\gamma} = 2b K_{\alpha\beta\gamma},
\end{equation}
where $b$ is a constant whose value gives the weight $2b$ of the Killing
tensor (cf. \cite{ru:kt}). At least in the case in which the metric
admits no Killing vector one can show that this is not a restriction
(cf. \cite{karlovini:killing}).  The Killing tensor can then be
factorized as
\begin{equation}\lbeq{Killingfactor}
  K_{\alpha\beta\gamma}=e^{2bt}\hat{K}_{\alpha\beta\gamma},\quad
  \mcl{L}_\xi \hat{K}_{\alpha\beta\gamma} = 0.
\end{equation}
The null variables $u$, $\bar{u}$ will in general not be the Killing
tensor adapted null variables $U$ and $\bar{U}$, so from the outset
the functions $S(u)$ and $R(\bar{u})$, as well as the functions
$Q(\bar{u})$ and $P(\bar{u})$ introduced for type IIA and IIB
respectively, has to be assumed arbitrary rather than taking the
standard values $0$ or $1$. However, it follows from
\refeq{Killingfactor} that $S$ and $R$ must have the exponential
dependence
\begin{equation}
  S(u) = S_0e^{2(b-3)u} = S_0e^{2(b-3)(t+x)}, \quad R(\bar{u}) =
  R_0e^{2(b-3)\bar{u}} = R_0e^{2(b-3)(t-x)},
\end{equation}
and that the trace vector potential $\Phi$ up to an irrelevant
additive constant must be of the form
\begin{equation}\lbeq{Phiform}
  \Phi = e^{2(b-1)t}\phi(x).
\end{equation}
Substituting this into eq.\ \refeq{Killingparam}, using
$K_u=2\Phi_{,u}=\Phi_{,t}+\Phi_{,x}$,
$K_{\bar{u}}=-2\Phi_{,\bar{u}}=-\Phi_{,t}+\Phi_{,x}$, thus gives us
the following general parametrization of
$\hat{K}_{\alpha\beta\gamma}$:
\begin{equation}\lbeq{homotetkillparam}
  \begin{split}
    \hat{K}_{000} &= -e^{-3t}R_0e^{-2(b-3)x}F^{3/2} \\
    \hat{K}_{111} &= -e^{-3t}S_0e^{2(b-3)x}F^{3/2} \\
    \hat{K}_{001} &=
    -e^{-3t}\textstyle\frac{1}{2}[\phi'+2(b-1)\phi]F^{-1/2} \\ 
    \hat{K}_{011} &=
    -e^{-3t}\textstyle\frac{1}{2}[\phi'-2(b-1)\phi]F^{-1/2}. 
  \end{split}
\end{equation}
At this point it is in place to note that the corresponding
parametrization in the second rank case in \cite{ru:kt} contains an
error. The components of $\hat{K}_{MN}$ given in eq.\ (A4) are all
missing a factor $e^{-2t}$. The consequence is that if $2b$ is to be
interpreted as the weight of the Killing tensor
$K_{MN}=e^{2bt}\hat{K}_{MN}$, one must in what follows eq.\ (A4)
substitute $b$ by $b-1$. Therefore the error affects only the
interpretation of $b$, not the results in \cite{ru:kt}.  Depending on
the type of the Killing tensor, we now proceed as follows.

\subsection*{Type I}

Here eq.\ \refeq{PhiKrelarb} together with $G=\mcl{K}_{,u\bar{u}}$
shows that one can assume without loss of generality that $\mcl{K}$ is
of the form
\begin{equation}
  \mcl{K}=e^{2t}k(x).
\end{equation}
Hence by substituting this into eq.\ \refeq{KcondIarb} one obtains a
third order nonlinear ODE for the function $k(x)$ as the final
condition. In terms of the functions
\begin{equation}
   \Gamma=\frac{1}{2}(S_0e^{2(b-3)x}+R_0e^{-2(b-3)x}),\quad
   \Sigma=\frac{1}{2}(S_0e^{2(b-3)x}-R_0e^{-2(b-3)x}),
\end{equation}
this condition reads
\begin{align}\lbeq{httIcond}
  &[3/4\Sigma\,k'' + b/2\Gamma\,k' + (b-3)\Sigma\,k]k''' +
  (2b-3)\Gamma\,(k'')^2\\ \nn
  &+ (2b^2-3b-3)\Sigma\,k'k'' + 2(2b-3)(b-4)\Gamma\,kk'' -
  2b\Gamma\,(k')^2\\ \nn
  &+ 4(-2b^2+2b+3)\Sigma\,kk' - 8(2b-3)(b-2)\Gamma\,k^2 = 0,
\end{align}
Note that when $b\neq 3$, there is no loss of generality in assuming
$|R_0|=|S_0|=1$ since one can make the translations $t\rarr t+t_0$,
$x\rarr x+x_0$ with $t_0=-\frac{1}{4(b-3)}\ln{\!|R_0S_0|}$,
$x_0=\frac{1}{4(b-3)}\ln{\!|R_0/S_0|}$, under which $S_0\rarr
\sgn{\!(S_0)}$, $R_0\rarr \sgn{\!(R_0)}$. With $b=-12/7$ and $R_0=-S_0$, an
example of a nontrivial solution to eq.\ \refeq{httIcond} reads
\begin{equation}
  \left\{ \begin{array}{l} k(x) =
      \frac{49}{4}\sinh{\!(6x/7)}[\cosh{\!(6x/7)}]^{4/3}  \\[5pt]
           F(x) = \sinh{\!(6x/7)}[\cosh{\!(6x/7)}]^{-2/3}     \\
         \end{array}\right.
\end{equation}

\subsection*{Type II}

As before we shall assume that $R(\bar{u})=0$, thus letting the case
$S(u)=0$ be obtained via the transformation $u\lrarr \bar{u}$ which
here is equivalent to $x\rarr -x$.

\subsection*{Type IIA}

Here the relation $\Phi_{,u}=Q(\bar{u})G$ implies that
$Q(\bar{u})=Q_0e^{2(b-2)\bar{u}}=Q_0e^{2(b-2)(t-x)}$. When setting
$\phi(x)=S_0{}^{-1}Q_0{}^2e^{-2(3b-7)x}\psi(x)$ and substituting eq.\
\refeq{Phiform} into eq.\ \refeq{PhicondIIAarb}, one obtains the
condition
\begin{align}\lbeq{psiODE}
  &[1/16(\psi')^2 - 1/2(b-3)\psi\psi' + (b-3)^2\psi^2 -
  1/2(b-1){}^2\psi]\psi'' - 5/12(b-3)(\psi')^3\\ \nn
  &+ [4(b-3)^2\psi + 1/2(2b-3){}^2](\psi')^2 - [12(b-3)^3\psi^2 +
  2(b-2)(3b-7){}^2\psi]\psi'\\ \nn
  &+ 32/3(b-3)^4\psi^3 + 16(b-3)(b-2)^2\psi^2 = 0.
\end{align}
To standardize the Killing tensor, one can e.g. set $S_0=1$ while
letting $Q_0$ determine the overall factor of $F(x)$. An example of a
nontrivial solution to eq.\ \refeq{psiODE} using this
standardization is given by $b=7/4$, $S_0=1$, $Q_0=25/3$ and
\begin{equation}
  \left\{ \begin{array}{l} \psi(x) = \frac{6}{125}(1+e^{-5x/3})^3
      \\[5pt]
           F(x) = e^{2x}(e^{3x/2}+e^{-x/6})^2   \\ \end{array}\right.
\end{equation}

\subsection*{Type IIB}

Here $\Phi_{,\bar{u}}=P(\bar{u})$ implies that
$P(\bar{u})=P_0e^{2(b-1)\bar{u}}=P_0e^{2(b-1)(t-x)}$.  Setting
$F(x)=S_0{}^{-1/2}P_0{}^{1/2}e^{-2(b-2)x}H(x)$ and substituting
$G=e^{2t}F(x)$ into eq. \refeq{quasiIIBarb}, one obtains the condition 
\begin{equation}\lbeq{HODE}
  -1/2(H^2-1)H'+1/3(b-3)H^3+(b-1)H=0.
\end{equation}
Analogous to type IIA, one can standardize the Killing tensor by
setting $S_0=1$ while letting the value of $P_0$ determine the overall
factor of $F(x)$. The general solution to eq.\ \refeq{HODE} can for
all values of $b$ be written down implicitly and for several values of
$b$ it is possible solve the algebraic equation for $H(x)$. Here we
merely give the simplest nontrivial solution, for which $b=0$,
$S_0=1$, $P_0=1$ and
\begin{equation}
  \left\{ \begin{array}{l} H(x) = e^{-2x}\pm\sqrt{e^{-4x}-1}  \\[5pt] 
           F(x) = e^{2x}(1\pm\sqrt{1-e^{4x}}).   \\
         \end{array}\right.
\end{equation}

\subsection*{Metrics with two exponential terms}

Of special interest is the physically relevant case where $F(x)$ is of
the form \cite{ujr:hh}
\begin{equation}
  F(x) = C_1e^{2mx}+C_2e^{2nx}.
\end{equation}
Making this ansatz and working through the integrability conditions
for all three types of third rank Killing tensors yields five
different solutions for which $m\neq n$ which are given in table
\ref{table:twoexp}.
\begin{table}[htbp]
  \begin{equation}
  \begin{tabular}{|c|c|c|c|c|} \hline
          & $m$   & $n$    & $b$   & Killing tensor type \\ 
    \hline
    (i)   & $3$   & $1/3$  & $1$   & IIA        \\ \hline
    (ii)  & $2$   & $1/2$  & $3/2$ & IIA        \\ \hline
    (iii) & $3/5$ & $-1/5$ & $9/5$ & IIA        \\ \hline
    (iv)  & $1/3$ & $-1/3$ & $2$   & I, IIA     \\ \hline
    (v)  & $i\sqrt{3}$ & $-i\sqrt{3}$ & $3$ & I \\ \hline
  \end{tabular}
  \end{equation}
\caption{Geometries of the exponential type $ds^2=2e^{2t}(C_1
  e^{2mx}+C_2 e^{2nx})(-dt^2+dx^2)$ admitting a third rank Killing
  tensor.}
\label{table:twoexp}
\end{table}
Unfortunately, only the trigonometric case (v) defines a new
integrable geometry, as the geometries corresponding to the cases
(i)-(iv) also admit at least one second rank Killing tensor
\cite{ru:kt}. In the cases (ii)-(iv), the existence of a third rank
Killing tensor, namely the Nijenhuis bracket \cite{sommers:nijenhuis}
of two independent second rank Killing tensors, could actually have
been predicted from the outset. In case (iv) the geometry has a
non-null Killing vector, so in this case there are a number of ways to
construct a reducible third rank Killing tensor, which in table
\ref{table:twoexp} is reflected by the fact that the Killing tensor
type can be both I and IIA. 

\section{Concluding remarks}
We have shown that the classification of third rank
(1+1)-dimensional Killing tensors given in this paper can be used to 
find new explicit integrable geometries. Some examples of such
geometries were given and   many more can be constructed by using our
results. Possible applications include inflationary models with a
scalar field, anisotropic cosmologies and stellar models. In all of
these cases the field equations can be formulated as geodesic
equations on a (1+1)-dimensional geometry \cite{ujr:hh}.

Unlike the case of second rank Killing tensors the separation of the
geodesic equations for the third rank case cannot be done by a
coordinate transformation on the configuration space. Instead it is
necessary to apply a separating transformation which involves the
entire phase space in a nontrivial way. The theory of such
transformations is not fully understood. However, there does exist a
recipe for finding separating variables \cite{sklyanin:sepvar}.

\newpage

\end{document}